# Performance Analysis of LS and LMMSE Channel Estimation Techniques for LTE Downlink Systems

Abdelhakim Khlifi[1] and Ridha Bouallegue[2]

[1]National Engineering School of Tunis, Tunisia
abdelhakim.khlifi@gmail.com
[2]Sup'Com, Tunisia
ridha.bouallegue@gmail.com

**ABSTRACT**

*The main purpose of this paper is to study the performance of two linear channel estimators for LTE Downlink systems, the Least Square Error (LSE) and the Linear Minimum Mean Square Error (LMMSE). As LTE is a MIMO-OFDM based system, a cyclic prefix is inserted at the beginning of each transmitted OFDM symbol in order to completely suppress both inter-carrier interference (ICI) and inter-symbol interference (ISI). Usually, the cyclic prefix is equal to or longer than the channel length but in some cases and because of some unforeseen channel behaviour, the cyclic prefix can be shorter. Therefore, we propose to study the performance of the two linear estimators under the effect of the channel length. Computer simulations show that, in the case where the cyclic prefix is equal to or longer than the channel length, LMMSE performs better than LSE but at the cost of computational complexity. In the other case, LMMSE continue to improve its performance only for low SNR values but it degrades for high SNR values in which LS shows better performance for LTE Downlink systems. MATLAB Monte – Carlo simulations are used to evaluate the performance of the studied estimators in terms of Mean Square Error (MSE) and Bit Error Rate (BER) for $2x2$ LTE Downlink systems.*

## 1. INTRODUCTION

In order to satisfy the exponentialgrowing demand of wireless multimedia services, a high speed data access is required. Therefore, various techniques have been proposed in recent years to achieve high system capacities. Among them,we interest to the multiple-input multiple-output (MIMO).The MIMO concept has attracted lot of attention in wireless communications due to its potential to increase the system capacity without extra bandwidth [1].Multipath propagation usually causes selective frequency channels. To combat the effect of frequency-selective fading, MIMO is associated with orthogonal frequency-division multiplexing (OFDM) technique. OFDM is a modulation technique which transforms frequency selective channel into a set of parallel flat fading channels.A cyclic prefix CP is added at the beginning of each OFDM symbol to eliminate ICI and ISI. Theinserted cyclic prefix is equal to or longer than to the channel[2].

The 3GPP Long Term Evolution (LTE) is defining the next generation radio access network. LTE Downlink systems adopt Orthogonal Frequency Division Multiple Access (OFDMA) and MIMO to provide up to 100 Mbps (assuming a 2x2 MIMO system with 20MHz bandwidth).

The performance of a MIMO-OFDM communication systemsignificantlydependsupon the channel estimation. Channel estimation techniques for MIMO-OFDM systems were carried out in many articles e.g. [3] [4]. However, in most of these research works, the CP length is assumed to be equal or longer than the maximum propagation delay of the channel. But in some cases and because of some unforeseen channel behaviour, the cyclic prefix can be shorter than channel length. In this case, both ICI and ISI will be introduced and this makes the task of channel estimation more difficult. Equalization techniques that could flexibly detect the signals in both cases in MIMO-OFDM systems are discussed in [8] [9].

In this paper, we will focus on the study of the performance of LS and LMMSE channel estimation techniques for LTE Downlink systems under the effect of the channel length. The performance evaluation of the two estimators for LTE systems was discussed in many articles e.g. [10] [11]

In the rest of the paper, we give an overview of LTE Downlink system in section II. A LTE MIMO-OFDM system model is given in section III. We discuss the two linear channel estimation techniques, LS and LMMSE in section IV with their simulation results for their performances under the effect channel length given in section V. Conclusion is given in the last section.

## 2. OVERVIEW OF LTE DOWNLINK SYSTEM

According to [5], the duration of one frame in LTE Downlink system is 10 ms.Each LTE radio frame is divided into 10 sub-frames of 1 ms. As described in Figure 1, each sub-frame is divided into two time slots, each with duration of 0.5 ms. Each time slot consists of either 7 or 6 OFDM symbols depending on the length of the CP (normal or extended). In LTE Downlink physical layer, 12 consecutive subcarriers are grouped into one Physical Resource Block (PRB). A PRB has the duration of 1 time slot. Within one PRB, there are 84resource elements(12 subcarriers × 7OFDM symbols) for normal CP or 72 resource elements (12subcarriers × 6OFDM symbols) for extended CP.

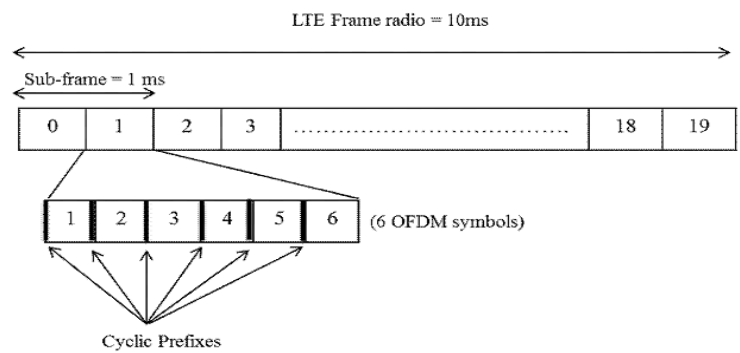

Figure 1: LTE radio Frame structure

LTE provides scalable bandwidth from 1.4 MHz to 20 MHz and supports both frequency division duplexing (FDD) and time-division duplexing (TDD). Table 1 shows the different transmission parameters for LTE Downlink systems.

Table 1: LTE Downlink parameters

| Transmission Bandwitdh (MHz) | 1.25 | 2.5 | 5 | 10 | 15 | 20 |
|---|---|---|---|---|---|---|
| Sub-frame duration (ms) | 0.5 | | | | | |
| Sub-carrier spacing (kHz) | 15 | | | | | |
| Sampling Frequency (MHz) | 1.92 | 3.84 | 7.68 | 15.36 | 23.04 | 30.72 |
| FFT size | 128 | 256 | 512 | 1024 | 1536 | 2048 |
| Number of occupied sub-carriers | 76 | 151 | 301 | 601 | 901 | 1201 |

## 3. DOWNLINK LTE SYSTEM MODEL

The system model is given in Figure.2. A MIMO-OFDM system with $N_{Tx}$ transmit and $N_{Rx}$ receive antennas is assumed.

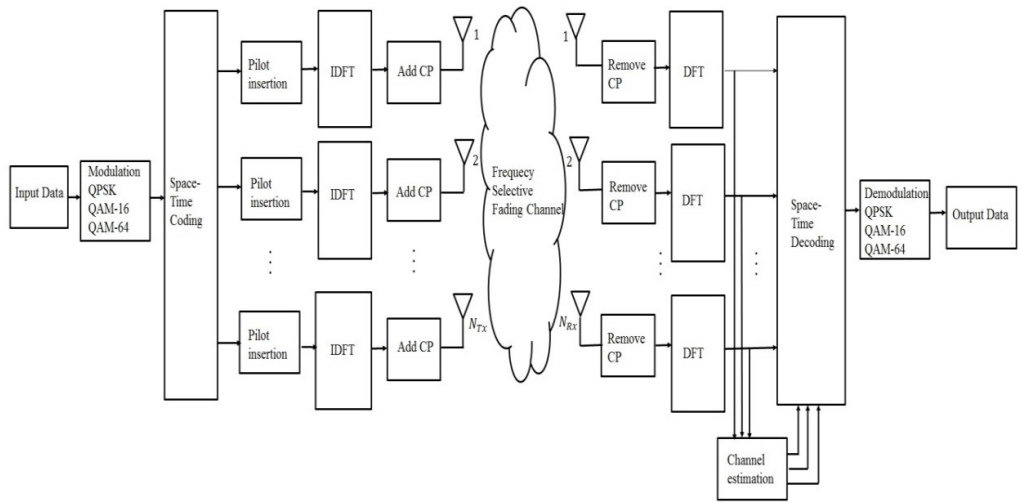

Figure 2: MIMO-OFDM system

OFDMA is employed as the multiplexing scheme in the LTE Downlink systems. OFDMA is a multiple users radio access technique based on OFDM technique. OFDM consists in dividing the transmission bandwidth into several orthogonal sub-carriers. The entire set of subcarriers is shared between different users.

Figure 3 illustrates a baseband OFDM system model. The $N$ complex constellation symbols $c_i$ are modulated on the orthogonal sub-carriers by mean of the Inverse Discrete Fourier. The $N$ subcarriers are spaced by $\Delta f = 15\ KHz$. To combat the effect of frequency-selective fading, a cyclic prefix (CP) with the length of $L_{CP}$ is inserted at the beginning of each OFDM symbol.

International Journal of Wireless & Mobile Networks (IJWMN) Vol. 3, No. 5, October 2011

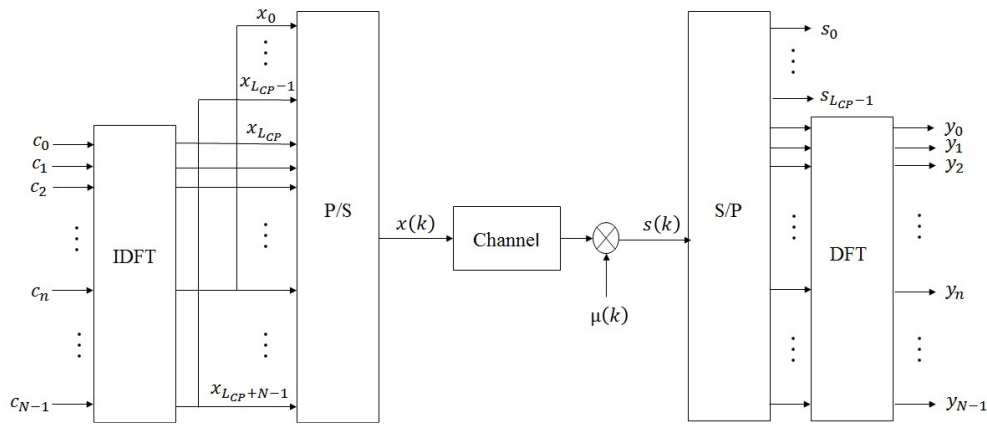

Figure 3: Baseband OFDM system

Each OFDM symbol is transmitted overfrequency-selective fading MIMO channels assumed independents of each other. Each channel ismodeled as a Finite Impulse Response (FIR) filter with $L$taps.
Therefore, we consider in our systemmodel only a single transmit and a single receive antenna. Afterremoving the CP and performing the DFT, the received OFDM symbol at one receive antenna can bewritten as:

$$Y = XH + \mu \quad (1)$$

$Y$ represents the received signal vector; $X$ is a matrix which contains the transmitted elements on its diagonal. $H$ is a channel frequency response, and $\mu$ is the noise vector whose entries have the i.i.d. complex Gaussian distribution with zero mean and variance$\sigma_\mu^2$.We assume that the noise $\mu$ is uncorrelated with the channel $H$.

## 4. CHANNEL ESTIMATION

In order to estimate the channel, LTE systems use pilot signals called reference signals.When short CP is used, they are being transmitted during the first and fifth OFDM symbols of every slot. When long CP is used, they are transmitted during the first and the fourth OFDM symbols.

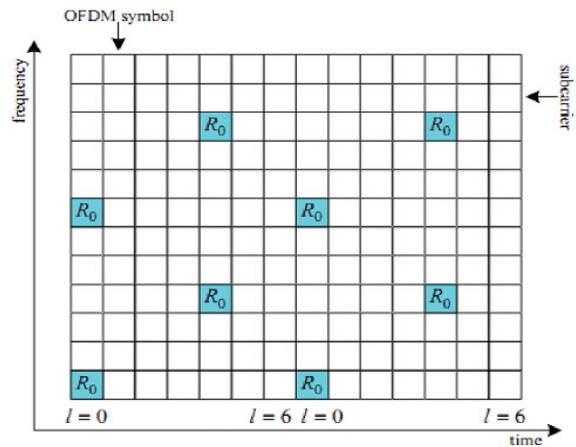

Figure 4: Downlink reference signal structure on one antenna port [5]

From (1), the received pilot signals can be written as:

$$Y_P = X_P H_P + \mu_P \qquad (2)$$

$(.)_P$ denotes positions where reference signals are transmitted.
In this paper, we study the performance ofLS and LMMSE channel estimation techniques.

## 4.1 Least Square LS

The goal of the channel least square estimator is to minimize the square distance between the received signal and the original signal.

The least square estimates (LS) of the channel at the pilot subcarriers given in (2) can be obtained by the following equation [6]:

$$\hat{H}_P^{LS} = (X_P)^{-1} Y_P \qquad (3)$$

$\hat{H}_P^{LS}$ represents the least-squares (LS) estimate obtained over the pilot subcarriers.

## 4.2 Linear Mean Minimum Square Error LMMSE

The LMMSE channel estimator is designed to minimize the estimation MSE.The LMMSE estimate of the channel responses given in (2) is [7]:

$$H_P^{LMMSE} = R_{HH_P}\bigl(R_{H_P H_P} + \sigma_\mu^2 (XX^H)^{-1}\bigr)\hat{H}_P^{LS} \qquad (4)$$

$R_{HH_P}$ represents the crosscorrelation matrix between all subcarriers and the subcarriers with reference signals.$R_{H_P H_P}$ represents the autocorrelation matrix of the subcarriers with reference signals.The high complexity of LMMSE estimator (4) is due to the inversion matrix lemma.Every time data changes, inversion is needed.The complexity of this estimator can be reduced by averaging the transmitted data. Therefore, we replace the term $(XX^H)^{-1}$ in (4) with its expectation $E[(XX^H)^{-1}]$.
The simplified LMMSE estimator becomes [7]:

$$\hat{H}_P^{LMMSE} = R_{HH_P}\left(R_{H_P H_P} + \frac{\beta}{SNR} I_P\right)^{-1} \hat{H}_P^{LS} \qquad (5)$$

where $\beta$ is scaling factor depending on the signal constellation (i.e $\beta = 1$ for QPSK and $\beta = 17/9$ for 16-QAM). SNR is the average signal-to-noise ratio, and $I_P$ is the identity matrix.

## 5. SIMULATION RESULTS

In this section, we compare the performance of the LS and the LMMSE estimation techniques for $2 \times 2$ LTE-5MHz Downlink system under the effect of the channel length. The transmitted signals are quadrature phase-shift keying (QPSK) modulated. The number of subcarriers in each OFDM symbol is $N = 300$, and the length of CP is $L_{CP}$=16. 100 LTE radio frames are sent through a frequency-selective channel. The frequency-selective fading channel responses are randomly generated with a Rayleigh probability distribution.

## 5.1 Case with $L \leq L_{CP}$

In this case, the cyclic prefix is longer than the channel which means that ISI and ICI are completely supressed. Figure 5 and Figure 6 shows that LMMSE estimation technique is better than the LS estimator.

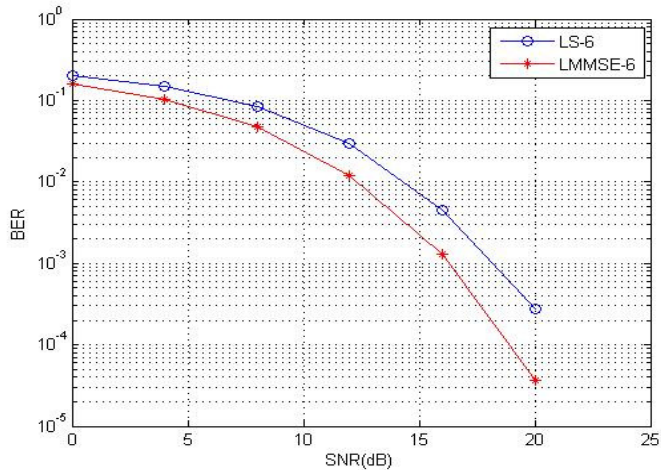

Figure 5: BER versus SNR for $L = 6$

Although, LMMSEgives the best performance but its complexity is higher due to the channel correlation and the matrix inversion lemma.

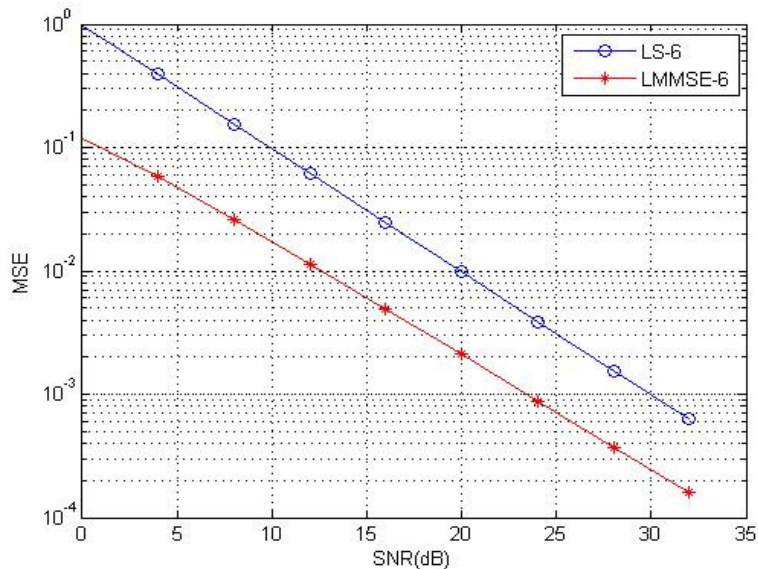

Figure 6: MSE versus SNR for $L = 6$

International Journal of Wireless & Mobile Networks (IJWMN) Vol. 3, No. 5, October 2011

## 5.2 Case with $L > L_{CP}$

In this case, the cyclic prefix is shorter than the channel. ISI and ICI will be introduced. Figure 7 shows that more the channel is longer than CP, more the performance is lost in terms of BER at the cost of more complexity for LMMSE estimation technique.

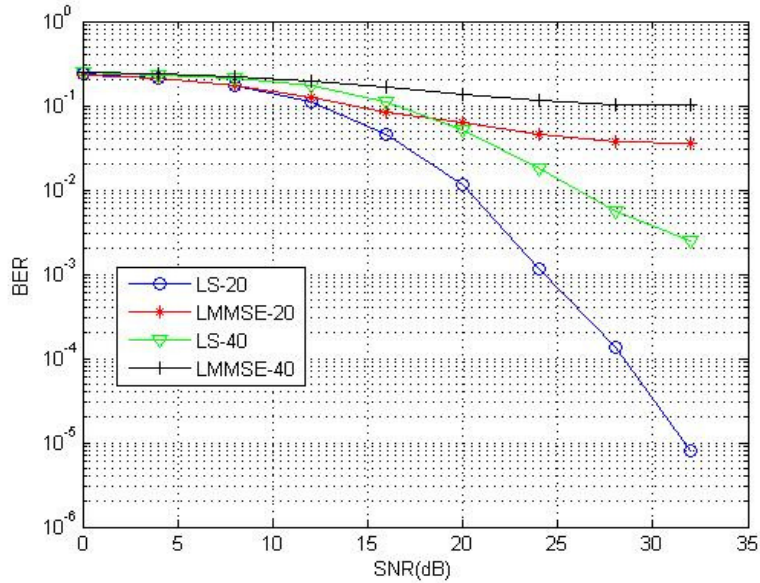

Figure 7: BER versus SNR for $L = 20$ and $L = 40$

Figure 8 demonstrates that LMMSE that; even the cyclic prefix is shorter than the channel length; LMMSE shows also better performances than LS for LTE Downlink systems but only for low SNR values. For high SNR values, LMMSE loses its performance in terms of MSE and LS estimator seems to perform better than LMMSE for this range of SNR values.

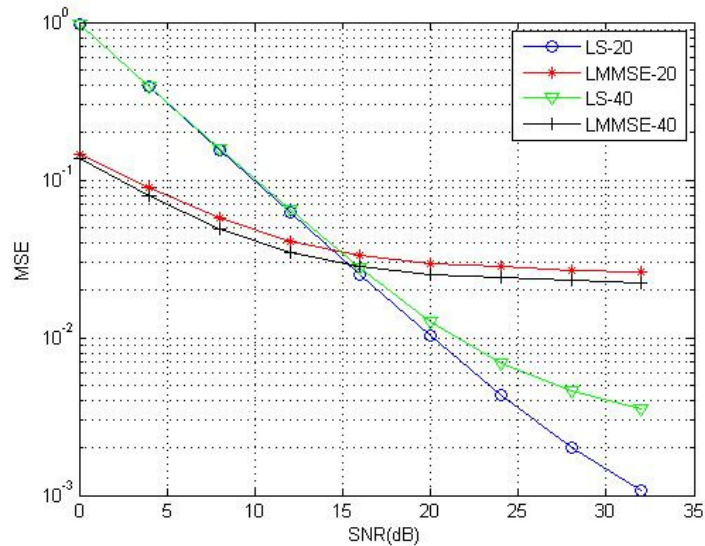

Figure 8: MSE versus SNR for $L = 20$ and $L = 40$

## 6. Conclusion

In this paper, we propose to evaluate the performance of LS and LMMSE estimation techniques for LTE Downlink systems under the effect of the channel length. The cyclic prefix inserted at the beginning of each OFDM symbol is usually equal to or longer than the channel length in order to suppress ICI and ISI. However, the CP length can be shorter than the channel length because of some unforeseen behaviour of the channel. Simulation results show that in the case where the CP length is equal to or longer than the channel length, the LMMSE performs better thanLS estimator but at the cost of the complexity because it depends on the channel and noise statistics. In the other case,LMMSEprovidesbetter performanceonly for low SNR values and begins to lose its performance for higher SNR values. In other hand, LS shows better performance than LMMSE in this range of SNR values.

**Authors**

**Abdelhakim KHLIFI** was born in Düsseldorf, Germany, on January 07, 1984. He graduated in Telecommunications Engineering, from National Engineering School of Gabès in Tunisia, July 2007. In June 2009, he received the master's degree of research in communication systems of the School of Engineering of Tunis ENIT. Currently he is a Ph.D student at the School of Engineering of Tunis. His research spans radio channel estimation in LTE MIMO-OFDM system.

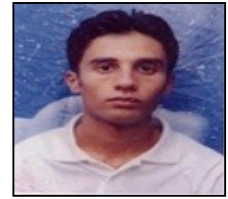

**Ridha BOUALLEGUE** is Professor at the National Engineering School of Tunis, Tunisia (ENIT). He practices at the Superior School of communication of Tunis (Sup'Com). He is founding in 2005 and Director of the Research Unit "Telecommunications Systems: 6'Tel@Sup'Comˇ T. He is founding in 2005, and Director of the National Engineering School of Sousse. He received his PhD in 1998 then HDR in 2003. His research and fundamental development, focus on the physical layer of telecommunication systems in particular on digital communications systems, MIMO, OFDM, CDMA, UWB, WiMAX, LTE, has published 2 book chapters, 75 articles in refereed conference lectures and 15 journal articles (2009).

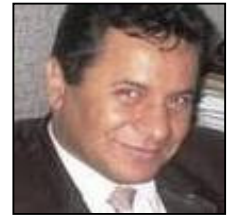